\documentclass[12pt]{article}
\usepackage[ascii]{inputenc}
\usepackage{amsmath,amssymb,amsfonts,amsthm}
\usepackage{mathtools}
\usepackage[margin=2cm]{geometry}
\usepackage{graphicx}
\usepackage{cite}
\usepackage{authblk}
\usepackage{color}
\usepackage[pdftex,bookmarks=false]{hyperref}
\usepackage{todonotes}
\DeclareMathAlphabet{\mathsfit}{T1}{\sfdefault}{\mddefault}{\sldefault}
\SetMathAlphabet{\mathsfit}{bold}{T1}{\sfdefault}{\bfdefault}{\sldefault}
\mathtoolsset{showonlyrefs}

\title{Equilibrium Spacetime Correlations of the Toda Lattice on the Hydrodynamic Scale}

\author{
Guido Mazzuca\footnote{Department of Mathematics, The Royal Institute of Technology, Stockholm, Sweden. \newline
\textit{Email: mazzuca@kth.se}}, 
Tamara Grava\footnote{
		International School for Advanced Studies (SISSA), Trieste, Italy,  School of Mathematics,  University of Bristol, UK  and INFN sezione di Trieste, \newline
		\textit{Email: grava@sissa.it} 
	},
Thomas Kriecherbauer\footnote{Department of Mathematics, Universit\"at Bayreuth, Germany\newline \textit{Email: thomas.kriecherbauer@uni-bayreuth.de} },
Kenneth T-R McLaughlin \footnote{Tulane University, New Orleans, United States \newline \textit{Email: kmclaughlin@tulane.edu }
},
Christian B.~Mendl\footnote{Technische Universit{\"a}t M{\"u}nchen
Department of Informatics, Boltzmannstra{\ss}e 3, 85748, Garching,
Germany \newline \textit{Email: christian.mendl@tum.de }} , 
Herbert Spohn\footnote{Technische Universit{\"a}t M{\"u}nchen
Department of Mathematics and Physics,
Boltzmannstra{\ss}e 3 and James-Franck-Str.~1, 85748 Garching,
Germany \newline \textit{Email: spohn@ma.tum.de}}}

\date{\today}

\begin{document}

\maketitle
\begin{abstract}
We report on molecular dynamics simulations of spacetime correlations of the Toda lattice in thermal equilibrium. The correlations of stretch, momentum, and energy are computed numerically over a wide range of pressure and temperature. Our numerical results are compared with the predictions from linearized generalized hydrodynamics on the Euler scale. The system size is $N=3000,4000$ and time $t =600$, at which ballistic scaling is well confirmed. With no adjustable parameters, the numerically obtained scaling functions agree with the theory within a precision of less than 3.5\%.
\end{abstract}

\section{Introduction}
\label{sec:intronew}
A central goal of Statistical Mechanics is to explore the structure of equilibrium correlations for observables of physical interest. 
These could be static correlations, but more ambitiously also correlations in spacetime. An interesting, but very fine-tuned class of hamiltonians are integrable many-body systems, either classical or quantum. This choice restricts us to systems
in one dimension. Then, generically, static correlations have exponential decay whether the model is integrable or not.
However, the dynamics of correlations is entirely different. In nonintegrable chains correlations propagate as 
a few narrow peaks at constant speed which then show characteristic sub-ballistic broadening. On the 
other hand for integrable models correlations still spread ballistically but now with a broad spectrum of velocities.
Such  behaviour was confirmed through a molecular dynamics (MD) simulation of the Ablowitz-Ladik model \cite{MS15}, an integrable discretization of  the nonlinear Schr\"{o}dinger equation. A further confirmation came from the simulation of the Toda chain \cite{KD16}. On the theoretical side, the 2016 construction of generalized hydrodynamics (GHD) was an important breakthrough \cite{BCDF16,CDY16}. This theory provides a powerful
tool through which, at least in principle, the precise form of the spectrum of correlations can be predicted. With such a development MD simulations can also be viewed as probing the validity of GHD. 

From the side of condensed matter physics, integrable quantum models have received considerable attention.  Because of size 
limitations, the simulation of macroscopic profiles are preferred. But time correlations have also been studied through
DMRG simulations \cite{BVKM17,BCM19, D19b, MP20}. In recent years,  attention has been given to the spacetime spin-spin correlation of the XXZ model at half-filling and at the isotropic point \cite{IDM18,LZP19,DP20}. The same quantity has also been investigated for a discrete classical chain
with 3-spins of unit length and interactions such that the model is integrable \cite{DKSD19}. A comparable situation occurs for the classical
sinh-Gordon equation, which is integrable as a nonlinear continuum wave equation and possesses  an integrable discretization,
see \cite{BDWY18} for MD simulations for equilibrium time correlations of the discrete model.  

In our contribution we study the correlations of the Toda chain in thermal equilibrium through MD simulations and compare with predictions from GHD. We will comment on the  connection to \cite{KD16} in the last section. To make our article reasonably self-contained we first discuss the Landau-Lifshitz theory for nonintegrable chains. This theory provides the connection between spacetime correlations and linearized hydrodynamics. For the Toda chain, the theory has to be extended so as to accommodate  an infinite number of conserved fields. We report on MD simulations of the Toda chain and compare with linearized GHD. 
 \section{Landau-Lifshitz theory}
\label{sec:intro}
The dynamics of the Toda chain is governed by the Hamiltonian
\begin{equation}\label{1}
  H=\sum_{j \in \mathbb{Z}}\big(\tfrac{1}{2}p^2_j +\exp(-(q_{j+1} - q_j))\big),
\end{equation}
where $(q_j,p_j)\in \mathbb{R}^2$ are position and momentum of the $j$-th particle \cite{T67,T89}. Introducing the $j$-th stretch (free volume) through $r_j = q_{j+1} - q_j$, the equations of motion read
\begin{equation}\label{2}
\frac{d}{dt}r_j=p_{j+1}-p_j\,,\qquad
\frac{d}{dt}p_j=- \mathrm{e}^{-r_j} + \mathrm{e}^{-r_{j-1}}, \qquad j \in\mathbb{Z}.
\end{equation}
By tradition, one introduces coefficients for the range
and strength of the interaction potential through $(g/\gamma)\exp(-\gamma(q_{j+1} - q_j))$. However, by a suitable change of spacetime scales, the form \eqref{2} can be regained, see the discussion in Section \ref{sec:outlook}. The Toda hamiltonian has no free parameters. Since the equilibrium measure for \eqref{1} is of product form, static correlations are easily accessible. 
Time correlations are more challenging, see \cite{SS80,S83}  
 for early attempts. 
A novel approach has been developed, now known as GHD. The guiding idea is to first identify the hydrodynamic equations for the Toda chain, which by necessity are a set of nonlinear
coupled hyperbolic conservation laws.  Given such an input one can construct the corresponding Landau-Lifshitz theory \cite{LL75,F75}, as
based on linearized GHD.    
  
Before entering into details, it will be useful to first recall the Landau-Lifshitz theory for a chain with a generic interaction potential, denoted by $V$ (for the Toda lattice $V(x) = \mathrm{e}^{-x}$), see \cite{S14} and references listed therein. Thus in \eqref{1} the interaction term reads $V(q_{j+1} - q_j)$ and the equations of motion become
\begin{equation}\label{3}
\frac{d}{dt}r_j=p_{j+1}-p_j\,,\qquad
\frac{d}{dt}p_j=V'(r_j)-V'(r_{j-1}).
\end{equation}
To define spacetime correlations we first have to specify the random initial data modelling thermal equilibrium. By Galileian invariance one restricts to the case of zero
average momentum. Then the Gibbs states are characterized by the inverse temperature $\beta >0$ and a parameter $P$ such that the physical pressure equals $P/\beta$.  For simplicity, we will also  refer to  $P$ as pressure. The allowed range of $P$
depends on $V$. If $V$ diverges faster than $|x|$ for $|x| \to \infty$, then $P\in \mathbb{R}$. For the Toda lattice $P>0$ because of the one-sided divergence of the exponential. In thermal equilibrium $\{(r_j,p_j), j\in\mathbb{Z}\}$ are a collection of i.i.d. random variables with single site probability density
\begin{equation}\label{4}
Z_0(P,\beta)^{-1} \exp\!\big(-\beta \big(\tfrac{1}{2}p_0^2 +V(r_0)\big) - Pr_0\big). 
\end{equation}
Here $Z_0(P,\beta)$ is the normalizing partition function. Note that, with our convention, $P$ and $\beta$ appear linearly in the exponent. Expectations with respect to such  i.i.d. random variables are denoted by $\langle \cdot \rangle_{P,\beta} $. We also shorten the notation for the covariance 
through $\langle X_1X_2 \rangle_{P,\beta}^\mathrm{c} = \langle X_1X_2 \rangle_{P,\beta} - \langle X_1 \rangle_{P,\beta} \langle X_2 \rangle_{P,\beta} $, where the particular random variables $X_1,X_2$ will be obvious from the context. 
 
For general $V$, the conserved fields are stretch, momentum, and energy with densities
\begin{equation}\label{5}
\vec{Q}(j)= \big(r_j, p_j,e_j\big), \qquad e_j=\tfrac{1}{2}p^2_j + V_j,
\end{equation}
using as  shorthand $V_j = V(r_j)$. $\vec{Q}$ is a three-vector with components labeled by $n=0,1,2$. The static space correlator is defined through
\begin{equation}\label{6}
\mathsfit{C}_{m,n}(j) =  \langle Q_m(j)Q_n(0)\rangle_{P,\beta}^\mathrm{c}
\end{equation}
and the static susceptibility by summing over space,
\begin{equation}\label{7}
\mathsfit{C}_{m,n} = \sum_{j \in \mathbb{Z}} \langle Q_m(j)Q_n(0)\rangle_{P,\beta}^\mathrm{c},
\end{equation}
$m,n = 0,1,2$. Since the underlying measure is product, only the  $j=0$ term is nonvanishing and
\begin{equation}\label{8}
\mathsfit{C} = \begin{pmatrix}
\langle r_0r_0\rangle_{P,\beta}^\mathrm{c} &0 & \langle r_0e_0 \rangle_{P,\beta}^\mathrm{c} \\[0.5ex]
0 & \langle p_0p_0\rangle_{P,\beta}^\mathrm{c} &0 \\[0.5ex]
\langle r_0e_0\rangle_{P,\beta}^\mathrm{c} &0& \langle e_0e_0\rangle_{P,\beta}^\mathrm{c}
\end{pmatrix},
\end{equation}
the zero entries resulting from $\langle p_0 \rangle_{P,\beta} =0$, $\langle p_0^3 \rangle_{P,\beta} =0 $, and $r_0,p_0$ being independent random variables.
 Later on we will need the statistics of the conserved fields on the hydrodynamic scale. More precisely, for smooth test functions $f$, we consider the random field
\begin{equation}\label{9}
\vec{\xi}_\epsilon(f) = \sqrt{\epsilon}\sum_{j\in \mathbb{Z}} f(\epsilon j) \big(\vec{Q}(j) - \langle\vec{Q}(0)\rangle_{P,\beta}\big). 
\end{equation}
Then, by the central limit theorem for independent random variables,
\begin{equation}\label{10}
\lim_{\epsilon \to 0}\vec{\xi}_\epsilon(f) = \int_\mathbb{R} \mathrm{d}x f(x) \vec{u}(x), 
\end{equation}
where the limit field $\vec{u}(x)$ is a Gaussian random field on $\mathbb{R}$ with mean zero, $\mathbb{E}(\vec{u}(x)) =0$,  and covariance 
\begin{equation}\label{11}
 \mathbb{E}(u_m(x)u_n(x')) = \mathsfit{C}_{m,n}\delta(x-x'),
\end{equation}
in other words, $\vec{u}(x)$ is Gaussian white noise with correlated components.

Microscopically, spacetime correlations are defined by evolving one of the observables to time $t$ which yields
\begin{equation}\label{12}
\mathsfit{S}_{m,n}(j,t) =  \langle Q_m(j,t)Q_n(0,0)\rangle_{P,\beta}^\mathrm{c}.
\end{equation}
Note that the Gibbs measure is spacetime stationary and thus without loss of generality both  arguments in $Q_{n}$ in \eqref{12} can be taken as $(0,0)$. To understand the structure of $\mathsfit{S}_{m,n}$ one has to rely on approximations. For the long time ballistic regime a standard scheme
is the Landau-Lifshitz theory, which views $Q_n(0,0)$ as a small perturbation of the initial Gibbs measure at the origin.  This perturbation will propagate and is then probed by the average of $Q_m$ at the spacetime point $(j,t)$. For large $(j,t)$ the microscopic dynamics is approximated
by the Euler equations, but only in their linearized version since the perturbation is small. More concretely, the approximate theory will be
a continuum field $\vec{u}(x,t)$ over $\mathbb{R}\times  \mathbb{R}$, which is governed by 
\begin{equation}\label{13}
\partial_t \vec{u}(x,t) + \mathsfit{A}\partial_x \vec{u}(x,t) = 0\,,
\end{equation}
with random initial conditions as specified in \eqref{11}.
The $3\times 3$ matrix $\mathsfit{A}$ is constant, i.e. independent of $(x,t)$. To explain the structure of $\mathsfit{A}$ requires some further efforts. We refer to \cite{S14} for more details and proofs of the key identities.

From the equations of motion one infers that to each density $Q_n(j,t)$ there is a current density  $J_n(j,t)$
such that 
\begin{equation}\label{14}
\frac{d}{dt}Q_n(j,t)  + J_n(j+1,t) - J_n(j,t)  = 0.
\end{equation}
Explicitly, the current densities are
\begin{equation}\label{15}
\vec{J}(j)  = -(p_j,V'_{j-1},p_jV'_{j-1}),
\end{equation}
where we adopted the convention  that omission of time argument $t$ means time $0$ fields. One then defines the static current-conserved field correlator
\begin{equation}\label{16}
\mathsfit{B}_{m,n}(j) =  \langle J_m(j)Q_n(0)\rangle_{P,\beta}^\mathrm{c},
\end{equation}
and the corresponding susceptibility
\begin{equation}\label{17}
\mathsfit{B}_{m,n} = \sum_{j \in \mathbb{Z}} \langle J_m(j)Q_n(0)\rangle_{P,\beta}^\mathrm{c}.
\end{equation}
Despite its asymmetric looking definition,
\begin{equation}\label{18}
\mathsfit{B}_{m,n} = \mathsfit{B}_{n,m}. 
\end{equation}

As a general property, Euler equations are built on thermally averaged currents. Linearizing them with respect to the average fields yields
\begin{equation}\label{19}
\mathsfit{A} = \mathsfit{B}\mathsfit{C}^{-1}. 
\end{equation}
Here $\mathsfit{B}$ appears when differentiating the average currents with respect to the chemical potentials and  $\mathsfit{C}^{-1}$ when switching from intensive to extensive variables.
By construction $\mathsfit{C} = \mathsfit{C}^\mathrm{T}$ and   $\mathsfit{C} >0$, in addition  $\mathsfit{B} = \mathsfit{B}^\mathrm{T}$ according to \eqref{18}. Hence
\begin{equation}\label{20}
\mathsfit{A} = \mathsfit{C}^{1/2} \mathsfit{C}^{-1/2} \mathsfit{B}\mathsfit{C}^{-1/2} \mathsfit{C}^{-1/2}, 
\end{equation}
which ensures that $\mathsfit{A}$ has real eigenvalues and a complete set of left-right eigenvectors. Anharmonic lattices 
are symmetric under time reversal, which implies the eigenvalues $\vec{c} = (-c,0,c)$, with $c>0$ the isentropic speed of sound.
We denote the right, resp. left eigenvectors of $\mathsfit{A}$ by $|\psi_\alpha\rangle$ and $\langle\tilde{\psi}_\alpha|$, $\alpha = 0,1,2$. With this input
the solution to \eqref{13} with initial conditions \eqref{11} reads
\begin{eqnarray}\label{21}
&&\hspace{-30pt}\mathsfit{S}^{\mathrm{LL}}_{m,n}(x,t) = \mathbb{E}\big(u_m(x,t)u_n(0,0)\big) \nonumber\\
&&\hspace{31pt}
= (\delta(x - \mathsfit{A} t)\mathsfit{C})_{m,n} = \sum_{\alpha = 0}^2 \delta(x - c_\alpha t) (|\psi_\alpha\rangle \langle \tilde{\psi}_\alpha | \mathsfit{C})_{m,n}
\end{eqnarray}
with $m,n = 0,1,2$. There are three $\delta$-peaks, the heat peak standing still and two sound peaks propagating in opposite directions with speed $c$.
Specifying $m,n$, each peak has a signed weight which depends on $\mathsfit{C}$ and the left-right eigenvectors of $\mathsfit{A}$.

The Landau-Lifshitz theory asserts that the microscopic correlator 
\begin{equation}\label{21a}
\mathsfit{S}_{m,n}(j,t) \simeq \mathsfit{S}^\mathrm{LL}_{m,n}(x,t)
\end{equation}
for $ j=\lfloor xt\rfloor$, $\lfloor \cdot\rfloor$ denoting integer part, with $t$ sufficiently large. The reader might be disappointed by the conclusion.  But with such basic information  the fine-structure of the peaks can be investigated, in particular their specific sub-ballistic broadening  and corresponding scaling functions \cite{S14,MS14,S17}. 

When turning to the Toda lattice, the conservation laws are now labeled by $n = 0,1, ...$ and thus $\mathsfit{A},\mathsfit{B},\mathsfit{C}$ become infinite dimensional matrices. The corresponding Landau-Lifshitz theory has been worked out in \cite{S20}. As to be discussed in the following section, with appropriate adjustments Eq. \eqref{21} is still valid.

\section{Toda lattice, linearized generalized hydrodynamics}  
\label{sec2}
The conservation laws of the Toda lattice are obtained from a Lax matrix \cite{F74,M75}. For this purpose, we first introduce the Flaschka variables 
\begin{equation}\label{22}
a_j = \mathrm{e}^{-r_j/2}.
\end{equation}
Then the equations of motion become
\begin{equation}\label{23}
\frac{d}{dt} a_j = \tfrac{1}{2}a_j(p_j - p_{j+1}),\quad \frac{d}{dt} p_j = a_{j-1}^2 - a_{j}^2.
\end{equation}
The Lax matrix, $L$, is defined by
\begin{equation}\label{24}
L_{j,j} = p_j,\qquad L_{j,j+1}= L_{j+1,j} = a_j,  
\end{equation}
$j \in \mathbb{Z}$, and $L_{i,j} = 0$ otherwise.  Clearly $L = L^\mathrm{T}$. The conserved fields are labelled by nonnegative integers
and have densities given by  
\begin{equation}\label{25}
Q_0(j) = r_j,\qquad Q_n(j) =  (L^n)_{j,j}\,,
\end{equation}
with $n\geq 1$. Note that $Q_n(j)$ is local in the sense that it depends only on the variables with indices in
the interval $[j-n,j+n]$.  An explicit expression for these quantities is given in \cite{GMMP20}. For the current densities one obtains
\begin{equation}\label{26}
J_0(j) = -p_j, \qquad J_n(j) =  (L^nL^{\scriptscriptstyle \downarrow})_{j,j}, \quad n=1,2,...\,,
\end{equation}
where $L^{\scriptscriptstyle \downarrow}$ is the lower triangular part of $L$. Then under the Toda dynamics 
\begin{equation}\label{27}
\frac{d}{dt} Q_n(j,t) + J_n(j+1,t) -J_n(j,t)  = 0,
\end{equation} 
which is the $n$-th conservation law in local form.

The first items in the list are stretch and momentum for which our current definitions agree with those in \eqref{5}, \eqref{15}. 
However, for $n=2$ one obtains $(L^2)_{0,0} = p_0^2 + a_{-1}^2 + a_0^2$ and  $(L^2L^{\scriptscriptstyle \downarrow})_{0,0}
= a_{-1}^2(p_{-1} + p_0)$, which differs from \eqref{5}, \eqref{15} on two accounts. First there is the trivial factor of $2$.
In our numerical plots we use the physical energy density $e_j$. The second point is more subtle. 
Densities are not uniquely defined, since one can add a difference of some local function and its shift by one.
When summing a particular choice for the density over some spatial interval,  the result differs from another choice of the density by a boundary term only.
Thus the bulk term  will have a correction of order 1/(length of interval), which does not affect the hydrodynamic equations. 
For the currents the difference can be written as a total time derivative which is again a boundary term when integrating 
over some time interval. In this section we adopt the conventions
\eqref{25} and \eqref{26}, since the analysis heavily relies on the Lax matrix. Beyond $n=2$, while the fields no longer have a name, they still have to be taken into account in a hydrodynamic theory. 

The infinite volume static field-field correlator is defined as in \eqref{6} and the 
current-field correlator as in  \eqref{16}. In particularly, $B = B^\mathrm{T}$. Of course, 
$C, B$ are now matrices in the Hilbert space of sequences indexed by $\mathbb{N}_0$, 
i.e. the space $\ell_2(\mathbb{N}_0)$. To distinguish $3\times 3$ matrices from their infinite dimensional counterparts,
for the latter we use standard italic symbols.
The spacetime correlator of the Toda lattice is defined by 
\begin{equation}\label{28}
S_{m,n}(j,t) =  \langle Q_m(j,t)Q_n(0,0)\rangle_{P,\beta}^\mathrm{c}.
\end{equation}
and we plan to construct its Landau-Litshitz approximation.
In essence this amounts to an analysis of
\begin{equation}\label{28a}
\big(\mathrm{e}^{At}C\big)_{m,n},\qquad A = BC^{-1}. 
\end{equation} 
While we are mainly interested in the physical fields corresponding to the indices $m,n=0,1,2$, for the operator in \eqref{28a} an understanding of the infinite dimensional matrices is required.

 Starting from the basics, the free energy of the Toda lattice is given by
\begin{equation}\label{29}
F_\mathrm{eq}(P,\beta) =  \log \sqrt{\beta/2\pi} +P\log \beta - \log\Gamma(P).
\end{equation}
In particular, the average stretch, $\nu$, is determined through
 \begin{equation}\label{30}
\nu(P,\beta) = \partial_P F_\mathrm{eq}(P,\beta) =  \langle Q_0(0)\rangle_{P,\beta}
=\log \beta -\psi(P),
\end{equation}
with $\psi$ the digamma function. Expectations of higher order fields can be written as moments of a probability measure denoted by $\nu\rho_\mathsf{p}$,
\begin{equation}\label{31} 
\kappa_n = \langle Q_n(0)\rangle_{P,\beta} = \int_\mathbb{R}\mathrm{d}w \nu\rho_\mathsf{p}(w)w^n, 
\end{equation} 
$n \geq 1$. $\rho_\mathsf{p}$ is called particle density. 
To determine this density one first has to solve the thermodynamic Bethe equations (TBA). For this purpose we introduce the integral operator
 \begin{equation}\label{32}
Tf(w) = 2 \int_\mathbb{R} \mathrm{d}w' \log |w-w'| f(w'),
\end{equation} 
$w \in \mathbb{R}$, considered as an operator on $L^2(\mathbb{R},\mathrm{d}w)$ and define the number density
\begin{equation}\label{33}
\rho_\mathsf{n}(w) = \mathrm{e}^{-\varepsilon(w)},
\end{equation} 
with quasi-energies $\varepsilon$.
The quasi-energies satisfy the TBA equation
\begin{equation}\label{34} 
\varepsilon(w) = \tfrac{1}{2}\beta w^2 -\mu - (T \mathrm{e}^{-\varepsilon})(w),
\end{equation}
where the chemical potential $\mu$ has to be adjusted such that 
\begin{equation}\label{35} 
\int _\mathbb{R}\mathrm{d}w  \rho_\mathsf{n}(w) =P.
 \end{equation}
Thereby the number density depends on the parameters $P$ and $\beta$.

The TBA equation is closely connected to the $\beta$-ensemble of random matrix 
theory. We rewrite
\eqref{34} as
\begin{equation}\label{34a} 
-\log\rho_\mathsf{n}(w) = \tfrac{1}{2}\alpha w^2 -\mu - \alpha P(T \rho_\mathsf{n})(w).
\end{equation}
As $\alpha \to \infty$, the entropy term on the lefthand side can be neglected and one recognizes the defining equation for the Wigner semi-cirle law on the interval $[-2\sqrt{P},2\sqrt{P}]$. The Lax DOS is the $P$-derivative of $\rho_\mathsf{n}$,
which diverges as $(w \pm 2\sqrt{P})^{-1/2}$ at the two borders.
As $\alpha$ is lowered the borders become smeared to eventually cross over to a Gaussian.

In practice, the TBA equation has to be solved numerically. But for thermal equilibrium
an exact solution is available \cite{O85,ABG12,FM2021}. Denoting the solution of \eqref{34} for $\beta=1$  and the constraint \eqref{35}  by 
$\rho_\mathsf{n}^*$ one has
 \begin{equation}\label{3.68}
\rho_\mathsf{n}^*(w) = \frac{\mathrm{e}^{-w^2/2}}{\sqrt{2 \pi}|\hat{f}_P(w)|^2}, \quad
\hat{f}_P(w) = 
\int_0^\infty \mathrm{d}t f_P(t) \mathrm{e}^{\mathrm{i}wt}, \quad f_P(t) = \sqrt{2}\pi^{-1}\Gamma(P)^{-1/2}t^{P - 1}  \mathrm{e}^{-\frac{1}{2}t^2}. 
\end{equation}
In our numerical simulations it is of advantage to use the exact solution.

The TBA equation is a standard tool from GHD as one way to write the Euler-Lagrange equations for the variational principle associated with the generalized free energy. For the Toda lattice such a variational formula was obtained in \cite{D19,S19}. Proofs using methods from the theory of large deviations and transfer operator method have also become available \cite{GM21,MM22,Mazzuca2022,M22}.

Next we introduce the dressing transformation of some function $f$ by
\begin{equation}\label{36} 
f^\mathrm{dr}= \big(1 - T\rho_\mathsf{n}\big)^{-1} f
\end{equation}
with $\rho_\mathsf{n}$ regarded as a multiplication operator. Then number and particle density are related as
\begin{equation}
\label{37} 
\rho_\mathsf{n}(w)= \frac{\rho_\mathsf{p}(w)}{1 +  T\rho_\mathsf{p}(w)} 
 \end{equation}
with inverse
\begin{equation}\label{38} 
\rho_\mathsf{p}= (1 - \rho_\mathsf{n} T)^{-1} \rho_\mathsf{n}  = \rho_\mathsf{n} \varsigma_0^\mathrm{dr}, 
 \end{equation}
using the convention $\varsigma_n(w) = w^n$.

For the average currents similar identities are available. The central novel quantity is the effective velocity
\begin{equation}\label{39} 
v^\mathrm{eff} = \frac{\varsigma_1^\mathrm{dr}}{\varsigma_0^\mathrm{dr}},
\end{equation}
see \cite{BCDF16,CDY16,S20a,YS20}. Then
\begin{equation}\label{40} 
\langle J_0(0)\rangle_{P,\beta} = -\kappa_1,
\end{equation} 
and, for $n\geq 1$,
\begin{equation}\label{41} 
\langle J_n(0)\rangle_{P,\beta} = 
\int_\mathbb{R}\mathrm{d}w \rho_\mathsf{p}(w)(v^\mathrm{eff}(w) -\kappa_1)w^n.
\end{equation} 
In thermal equilibrium we have $\kappa_1=0$.

Since in the following there will be many integrals over $\mathbb{R}$, let us first
introduce the abbreviation
\begin{equation}\label{42} 
\langle f \rangle = 
\int_\mathbb{R}\mathrm{d}w f(w).
\end{equation}
With this  notation the $C$ matrix turns out to be of the form
\begin{eqnarray}\label{43}
&&\hspace{-20pt} C_{0,0} =\nu^3 \langle \rho_\mathsf{p} \varsigma_0^\mathrm{dr} \varsigma_0^\mathrm{dr} \rangle,\nonumber\\[0.5ex]
&&\hspace{-20pt}C_{0,n}= C_{n,0}= -\nu^2  \langle \rho_\mathsf{p} \varsigma_0^\mathrm{dr} (\varsigma_n- \kappa_n\varsigma_0)^\mathrm{dr} \rangle,\nonumber\\[0.5ex]
&&\hspace{-20pt} C_{m,n}= \nu  \langle \rho_\mathsf{p} (\varsigma_m- \kappa_m\varsigma_0)^\mathrm{dr} (\varsigma_n- \kappa_n\varsigma_0)^\mathrm{dr} \rangle,
\end{eqnarray}
 $m,n \geq 1$.  Note that the matrix $C$ has the block structure
 \begin{eqnarray}\label{44} 
C = \begin{pmatrix}
C_{0,0} & C_{0,n}\\
C_{m,0} & C_{m,n}
\end{pmatrix},
\end{eqnarray} 
 in the sense that $C_{m,n}$ for $m,n \geq 1$ follows a simple pattern. This structure will reappear for $B$ and 
 $\mathrm{e}^{At}C$.
 
 The field-current correlator $B$ can be computed in a similar fashion with the result 
\begin{eqnarray}\label{45}
&&\hspace{-20pt} B_{0,0} = \nu^2 \langle \rho_\mathsf{p}(v^\mathrm{eff} -\kappa_1) \varsigma_0^\mathrm{dr}   \varsigma_0^\mathrm{dr}\rangle,\nonumber\\[0.5ex]
&&\hspace{-20pt}B_{0,n}= B_{n,0}=  -\nu \langle \rho_\mathsf{p}(v^\mathrm{eff} -\kappa_1)  \varsigma_0^\mathrm{dr} ( \varsigma_n- \kappa_n \varsigma_0)^\mathrm{dr}\rangle,\nonumber\\[0.5ex]
&&\hspace{-20pt} B_{m,n} =  \langle \rho_\mathsf{p}(v^\mathrm{eff} -\kappa_1)( \varsigma_m- \kappa_m \varsigma_0)^\mathrm{dr} 
( \varsigma_n- \kappa_n \varsigma_0)^\mathrm{dr}\rangle.
\end{eqnarray}

As in \eqref{21}, we want to determine the propagator of the Landau-Lifshitz theory, denoted by $S^\mathrm{LL}_{m,n}(x,t)$. 
In principle, all pieces have been assembled. However to figure out the exponential of $A$ requires
its diagonalization. Details can be found in \cite{S20} and we only mention that one constructs a linear similarity transformation, $R$, such that
$R^{-1}AR$  is multiplication by
\begin{equation}\label{46} 
\nu^{-1}(v^\mathrm{eff}(w) -\kappa_1)
\end{equation}
in $L^2(\mathbb{R}, \mathrm{d}w)$.
Here $v^\mathrm{eff}$ is the effective velocity defined in \eqref{39}. Using the block convention as in \eqref{44}, the spacetime 
correlator in the Landau-Lifshitz approximation is given by
 \begin{eqnarray}\label{47}
&&\hspace{-40pt}S^\mathrm{LL}(x,t) =  \int_\mathbb{R}\mathrm{d}w   \delta\big(x- t\nu^{-1}(v^\mathrm{eff}(w) - \kappa_1)\big) \nu\rho_\mathsf{p}(w)\\[0.5ex]
&&\times\begin{pmatrix}
\nu^2\varsigma_0^\mathrm{dr}(w)^2 & \nu\varsigma_0^\mathrm{dr}(w)(\varsigma_n- \kappa_n \varsigma_0)^\mathrm{dr}(w)\\[1ex]
\nu\varsigma_0^\mathrm{dr}(w)(\varsigma_m- \kappa_m \varsigma_0)^\mathrm{dr}(w) &
 (\varsigma_m- \kappa_m \varsigma_0)^\mathrm{dr}(w)( \varsigma_n- \kappa_n \varsigma_0)^\mathrm{dr}(w)\nonumber
\end{pmatrix}.
\end{eqnarray}
Note that $S^\mathrm{LL}(x,0) = \delta(x)C$. As a property of the Euler equations, the expression \eqref{47} possesses exact ballistic scaling, 
\begin{equation}\label{48} 
S^\mathrm{LL}_{m,n}(x,t) = \frac{1}{t} S^\mathrm{LL}_{m,n}(x/t,1).
\end{equation}
The correlator $S_{m,n}(j,t)$ is computed in our MD simulations  which will then be compared with $S^\mathrm{LL}_{m,n}(x,t)$.

\section{Numerical simulations}
\label{sec:numerics}

For a molecular dynamics simulation one has to first specify a finite ring $[1, \dots, N]$ with suitable boundary conditions. For the dynamics of positions $q_j$ and momenta $p_j$ one imposes
\begin{equation}\label{3.1}
q_{N+1} = q_1 + \nu N.
\end{equation}
The parameter $\nu$ fixes the free volume per particle and can have either sign. In our simulation, we actually allow for 
a fluctuating free volume by choosing random initial conditions such that $\{r_1, p_1, \dots, r_N, p_N\}$ are i.i.d. random variables with a single site distribution as specified in \eqref{4}. Then the deterministic time evolution is governed by \eqref{23} with boundary conditions
\begin{equation}
r_0 = r_N,\qquad p_{N+1} = p_1.
\end{equation}
In fact, the boundary condition in \eqref{3.1} amounts to the micro-canonical constraint
\begin{equation}
\sum_{j=1}^N r_j = \nu N.
\end{equation}
If  one sets $\nu = \langle Q_0(0)\rangle_{P,\beta}$, then for large $N$, by the equivalence of ensembles, the
two schemes for sampling the correlator $S_{m,n}(j,t)$ should differ by the size of statistical fluctuations. 
For a few representative examples we checked that indeed the equivalence of ensembles holds for the particular observables under study.

Returning to the choice of system size there is an important physical constraint.
In all simulations one observes a sharp right and left front, which travel with constant speed and beyond which spatial  correlations are exponentially small. On a ring necessarily the two fronts will collide after some time. Such an encounter has a noticeable effect on the molecular dynamics which is not captured by the linearized GHD analysis. Therefore  the simulation time is limited by the time of first collision. Indeed, we note  in Figures~\ref{fig:ontop_beta05}-\ref{fig:ontop_beta20} that both linearized GHD and MD clearly display maximal speeds of at most $\Delta j/\Delta t = 2$ for the entire range of $(P, \beta, m, n)$ displayed in these figures. 
Taking into account that the initial correlations are proportional to $\delta_{0j}$, we conclude that for a ring of size $N=3000$ there will be no collision of the two fronts up to time $t=750$ which is larger than $t=600$ used in our simulations. 

Before displaying and discussing our results, we provide more details on numerically solving the TBA equations 
and on the actual scheme used for MD.



\subsection{Details of the numerical implementation}

\subsubsection{Solving linearized GHD}
\label{sec:linearized_GHD}

To numerically solve the linearized GHD equations, we use a numerical method similar to the one from \cite{Mendl2022}. First, Eq.~\eqref{3.68} can be expressed in terms of the parabolic cylinder function $D_\nu(z)$, which is readily available in \texttt{Mathematica}. This provides the solution to the TBA equations \eqref{34}, \eqref{35}.


Then, we use a simple finite element discretization of the $w$-dependent functions by hat functions, resulting in piecewise linear functions on a uniform grid. After precomputing the integral operator $T$ in \eqref{33} for such hat functions, the dressing transformation \eqref{37} becomes a linear system of equations, which can be solved numerically. This procedure yields $\varsigma_n^\mathrm{dr}$, and subsequently $\rho_\mathsf{p}$ via \eqref{38} and $v^\mathrm{eff}$ via \eqref{39}. The moments can be computed from $\kappa_n = \int_\mathbb{R}\mathrm{d}w \nu \rho_\mathsf{n}(w) \varsigma_n^\mathrm{dr}(w)$, or (equivalently) Eq.~\eqref{31}.

To evaluate the correlator in \eqref{47}, we note that the delta-function in the integrand results in a parametrized curve, with the first coordinate (corresponding to $x/t$) equal to $\tilde{v}^\mathrm{eff}$ from \eqref{46}, and the second coordinate equal to the remaining terms in the integrand divided by the Jacobi factor $|\frac{\mathrm{d}}{\mathrm{d}w} \tilde{v}^\mathrm{eff}(w)|$ resulting from the delta-function.

\subsubsection{Molecular dynamics simulations}
\label{sec:md_simulations}

We approximate the expectation value that is contained in the MD-definition of the correlations $S_{m,n}$ in equation~\eqref{28} by the following numerical scheme, whose implementation program is written in \texttt{Python}, and can be found at \cite{Toda_repo}.
First, we generate the random initial conditions distributed according to the Gibbs measure, as given by \eqref{4} for the i.i.d.~random variables $(r_j, p_j)_{1\leq j \leq N}$. Specifically, the variables $p_j$ are distributed according to a standard normal random variable, that we generate with \texttt{Numpy v1.23}'s native function \texttt{random.default{\textunderscore}rng().normal} \cite{Numpy}, times $1/\sqrt{\beta}$. It takes a brief calculation to see that $r_j$ can be chosen to be $-\ln (X/(2\beta))$ where $X$ is chi-square distributed with shape parameter $2P$. We obtain the random variable $X$ using \texttt{Numpy v1.23}'s native function \texttt{random.default{\textunderscore}rng().chisquare}.
Having chosen the initial conditions in such a manner, we solve equation \eqref{2}.
For the evolution, we adapt the classical St\"ormer--Verlet algorithm \cite{Hairer2006} of order 2 to work with the variables $(\mathbf{p},\mathbf{r})$.  
Specifically, we used a time step equal to $\delta = 0.05$, and, given the solution $(\mathbf{r}(t),\mathbf{p}(t))$ at time $t$, we approximate the solution at time $t+\delta$ through the following scheme,

\begin{align}
    & p_j\left(t+\frac{\delta}{2}\right) = p_j(t) - \frac{\delta}{2}\left(e^{-r_j(t)} - e^{r_{j-1}(t)} \right) \,,\\
    &r_j(t+\delta) = r_j(t) + \delta\left(p_{j+1}\left(t+\frac{\delta}{2}\right) - p_j\left(t+ \frac{\delta}{2}\right)\right)\,,\\
    & p_j(t+\delta) = p_j\left( t+\frac{\delta}{2}\right) - \frac{\delta}{2}\left(e^{-r_j(t+\delta)} - e^{r_{j-1}(t+\delta)} \right)\,,
\end{align}
for all $j=1,\ldots,N$.
In this part of the implementation, we extensively used the library \texttt{Numba} \cite{numba} to speed up the computations. 


Our approximation for the expectation $S_{m,n}$ is then extracted from $3\times10^6$ trials with independent initial conditions. Here we take the empirical mean of all trials where for each trial we also take the mean of the $N=3000$ sets of data that are generated by choosing each site on the ring for $j=0$.

To evaluate
the quality of our numerical simulations, we have repeated the numerical experiments up to five times including variations for the length of the ring and evaluating the solutions at more intermediate time steps than displayed in the figures below. Furthermore, we have compared the results with the corresponding outcomes obtained by a \texttt{MATLAB} program that has been developed independently from the \texttt{Python} program, and that follows a different numerical scheme. It uses \texttt{MATLAB}'s random number generators \texttt{randn} for initial momenta and \texttt{rand} combined with the rejection method to produce initial stretches. The dynamics is then evaluated by the solver \texttt{ode45}, which exploits the Runge--Kutta method to numerically solve the Hamiltonian system associated with \eqref{1} on the ring. We found that the deviations between different experiments are comparable to the size of the amplitudes of the high frequency oscillations that are present in figures~\ref{fig:compa_error}-\ref{fig:compa_error2}. These oscillations are due to the random fluctuations of the empirical means around their expectation values $S_{m,n}$. Agreement of different experiments up to the order of these oscillations therefore shows the consistency of the corresponding numerical results.    

We also want to mention that all the pictures that appeared in this paper are made using the library \texttt{matplotlib} \cite{Matplotlib}.

\subsection{Comparison of linearized GHD with MD at time $t=600$}
\label{sec:comparisons}

\begin{figure}
\centering
\includegraphics[width=0.8\textwidth]{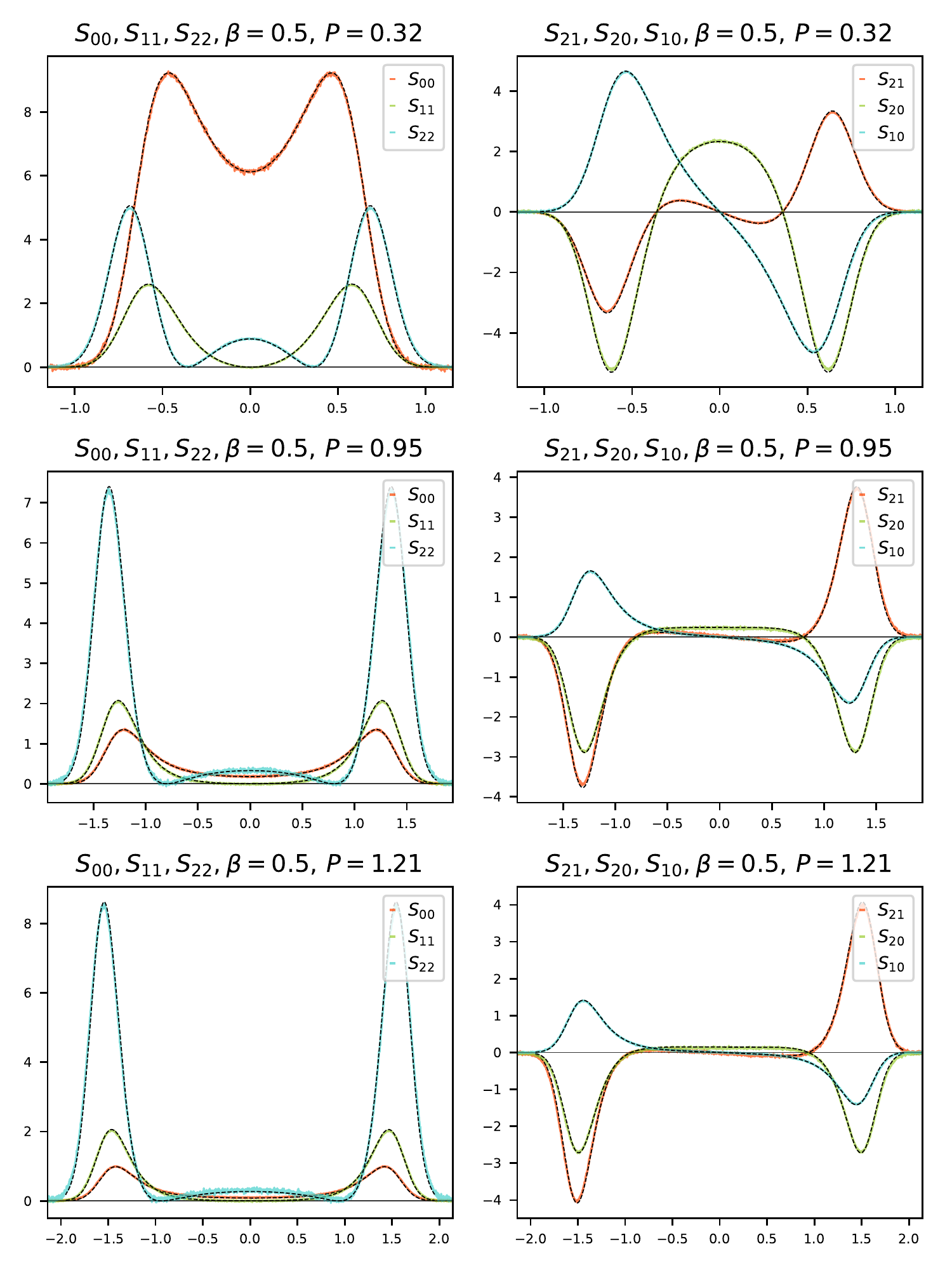}
\caption{Toda correlation functions: GHD predictions $y \mapsto S^\mathrm{LL}_{m,n}(y,1)$ vs.\ numerical simulations of the molecular dynamics $y \mapsto t S_{m,n}(yt,t)$ at $t = 600$ for $\beta=0.5$ with low pressure (top), medium pressure (middle) and high pressure (bottom). Numerical simulations are colored according to the legend, the corresponding GHD predictions are displayed by dashed lines. Number of trials: $3\times10^6$. }
\label{fig:ontop_beta05}
\end{figure}

\begin{figure}
\centering
\includegraphics[width=0.8\textwidth]{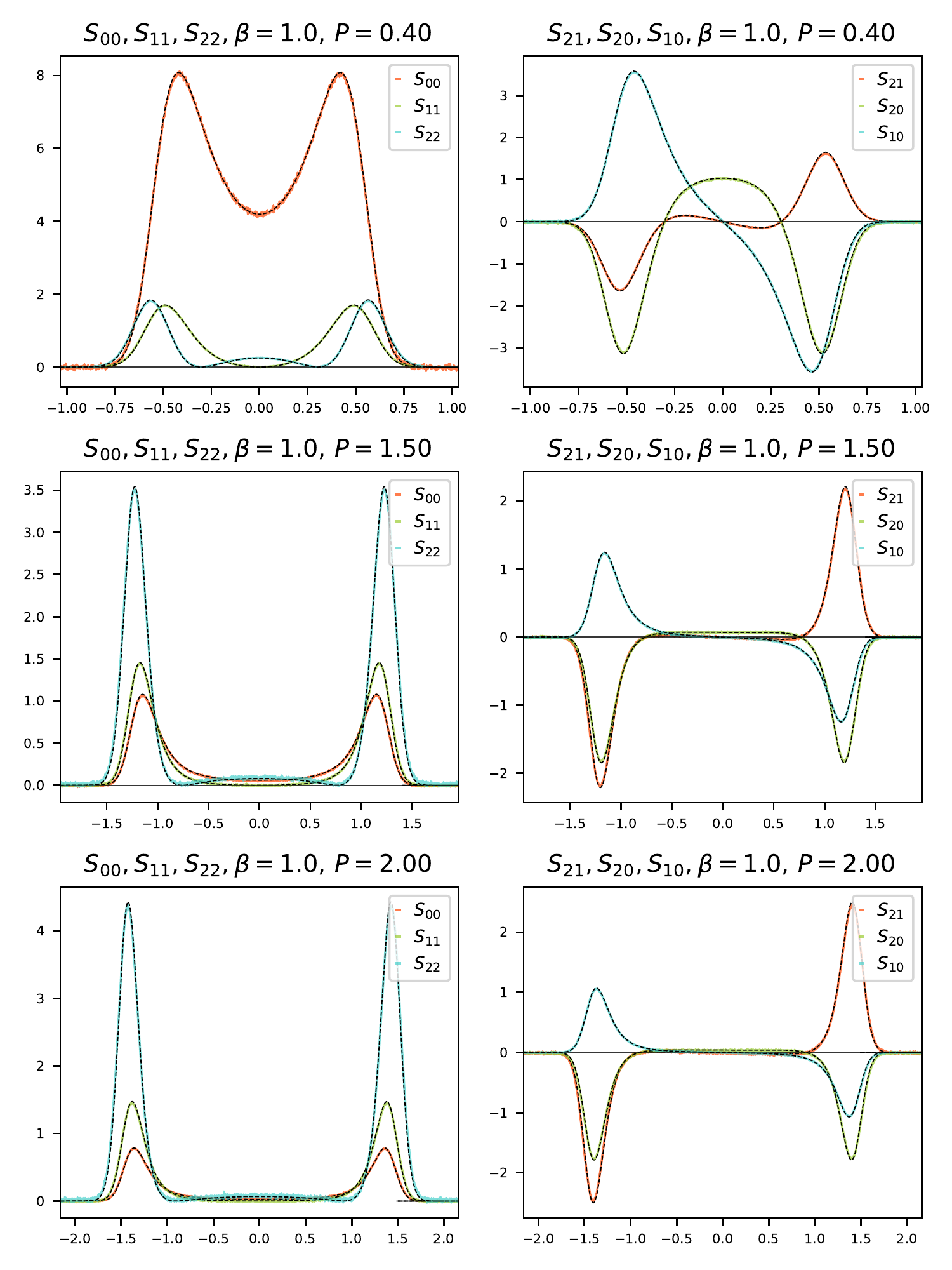}
\caption{Toda correlation functions: GHD predictions $y \mapsto S^\mathrm{LL}_{m,n}(y,1)$ vs.\ numerical simulations of the molecular dynamics $y \mapsto t S_{m,n}(yt,t)$ at $t = 600$ for $\beta=1.0$ with low pressure (top), medium pressure (middle) and high pressure (bottom). Numerical simulations are colored according to the legend, the corresponding GHD predictions are displayed by dashed lines. Number of trials: $3\times10^6$. }
\label{fig:ontop_beta10}
\end{figure}

\begin{figure}
\centering
\includegraphics[width=0.8\textwidth]{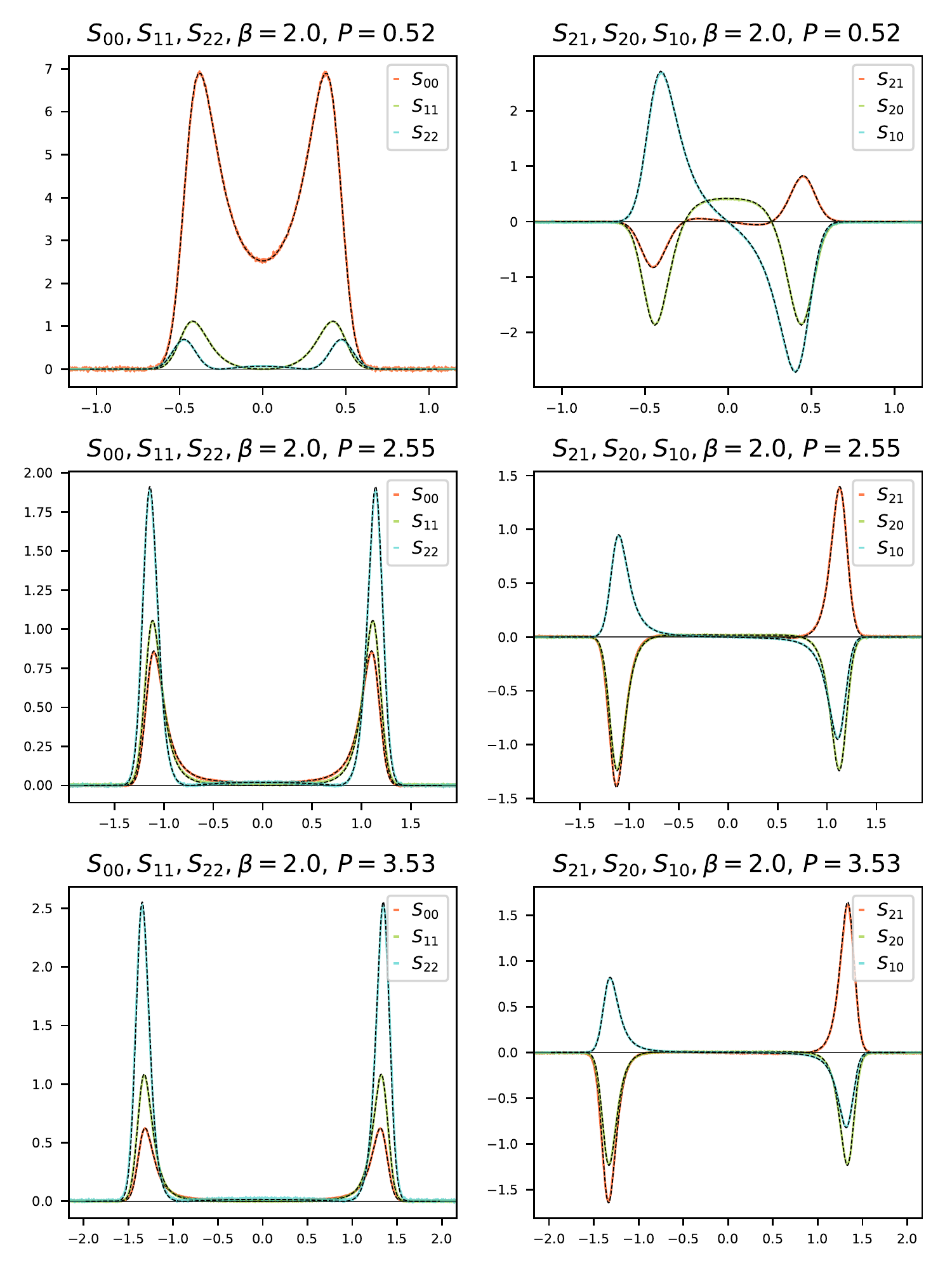}
\caption{Toda correlation functions: GHD predictions $y \mapsto S^\mathrm{LL}_{m,n}(y,1)$ vs.\ numerical simulations of the molecular dynamics $y \mapsto t S_{m,n}(yt,t)$ at $t = 600$ for $\beta=2.0$ with low pressure (top), medium pressure (middle) and high pressure (bottom). Numerical simulations are colored according to the legend, the corresponding GHD predictions are displayed by dashed lines. Number of trials: $3\times10^6$. }
\label{fig:ontop_beta20}
\end{figure}


We compare the GHD predictions with MD simulations for three different temperatures that correspond to $\beta = 0.5$ (Fig.~\ref{fig:ontop_beta05}), $\beta = 1$ (Fig.~\ref{fig:ontop_beta10}), and $\beta = 2$ (Fig.~\ref{fig:ontop_beta20}). For each $\beta$ we choose three different values for the pressure parameter $P$ in such a way that the corresponding mean stretches, given by \eqref{30}, are positive ($\approx 2.57$) for low pressure, negative ($\approx -0.42$) for high pressure and approximately zero for medium pressure. We summarize their values in Table~\ref{Table_average_stretches}.

\begin{table}[h]
\centering
\renewcommand{\arraystretch}{1.5}
\begin{tabular}{|l|c|c|c|} \hline
pressure & $\beta=0.5$&$\beta=1$&$\beta=2$ \\ \hline
low & $P=0.32, \;\langle r \rangle\approx+2.58$ & $P=0.4, \;\langle r \rangle\approx+2.56$ & $P=0.52, \;\langle r \rangle\approx+2.56$\\ \hline
medium & $P=0.95, \;\langle r \rangle\approx-0.03$ & $P=1.5, \;\langle r \rangle\approx-0.04$ & $P=2.55, \;\langle r \rangle\approx-0.03$ \\ \hline
high & $P=1.21, \;\langle r \rangle\approx-0.42$ & $P=2.0, \;\langle r \rangle\approx-0.42$ & $P=3.53, \;\langle r \rangle\approx-0.42$ \\ \hline
\end{tabular}
\caption{Values for $\beta$ and $P$ and the corresponding mean stretches used in experiments}
\label{Table_average_stretches}
\renewcommand{\arraystretch}{1}
\end{table}

In each of the nine cases we have evaluated the Landau-Lifshitz approximations $S^\mathrm{LL}_{m,n}(\cdot,1)$, see \eqref{47}, of the correlators for all $0\leq n\leq m\leq 2$ using the numerical scheme described in Section~\ref{sec:linearized_GHD}. Their graphs are displayed in Figures~\ref{fig:ontop_beta05}-\ref{fig:ontop_beta20} as dashed lines. Note that the speeds of the sound peaks depend significantly on both pressure and temperature. Moreover, the predicted fine-structure of both the heat and the sound peaks are quite different for low pressure when compared to medium and high pressure.

The colored lines in Figures~\ref{fig:ontop_beta05}-\ref{fig:ontop_beta20} show our numerical results for the corresponding molecular dynamics. According to the predicted ballistic scaling \eqref{48} we plot $t S_{m,n}(j,t)$ as a function of $j/t$ for $t=600$. Here the values of $S_{m,n}(j,t)$ are approximated using the numerics explained in Section~\ref{sec:md_simulations}.

The agreement between linearized GHD and MD is striking, in particular since there are no adjustable parameters. In all of the 54 comparisons shown in Figures~\ref{fig:ontop_beta05}-\ref{fig:ontop_beta20} the GHD predictions for the fine-structure of heat and sound peaks are in excellent agreement with the ones observed from molecular dynamics at time $t=600$. As we show in more detail in the next subsection the largest deviations occur mostly near the sound peaks and do not exceed $3.5\%$ of the peaks' maximal values.
\subsection{Deviation of linearized GHD from MD at times $t=150$ and $t=600$}
\label{sec:convergence}

\begin{figure}
\centering
\includegraphics{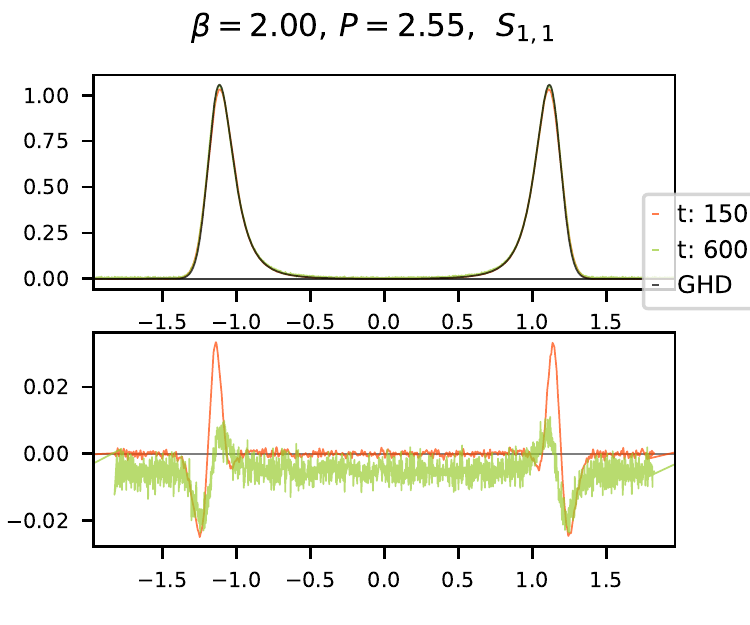}
\includegraphics{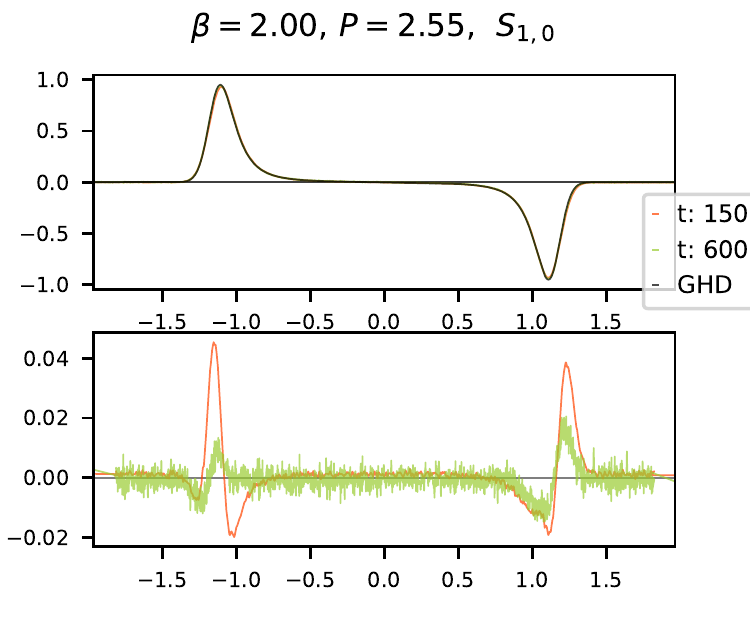}\\
\includegraphics{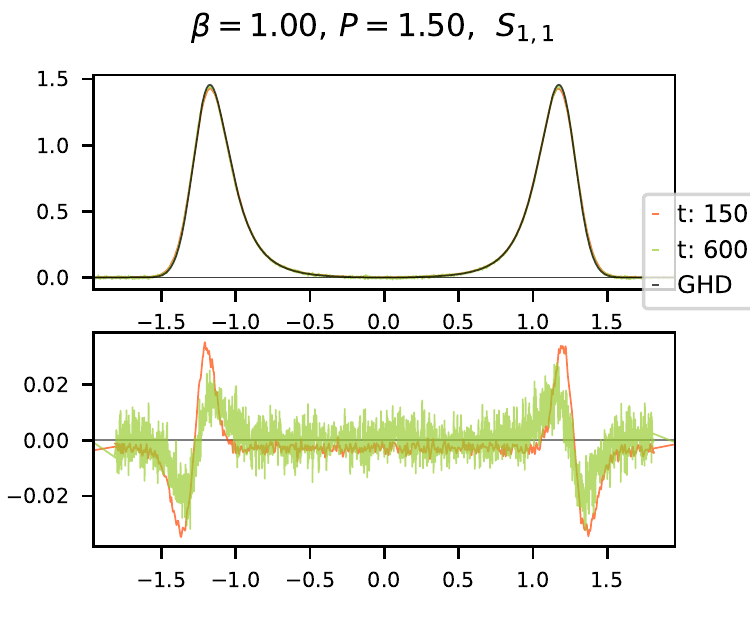}
\includegraphics{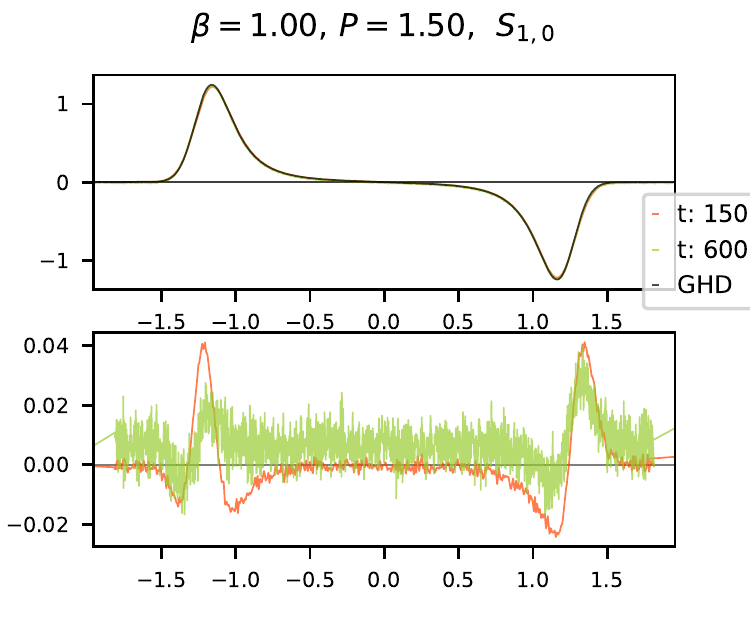} \\
\includegraphics{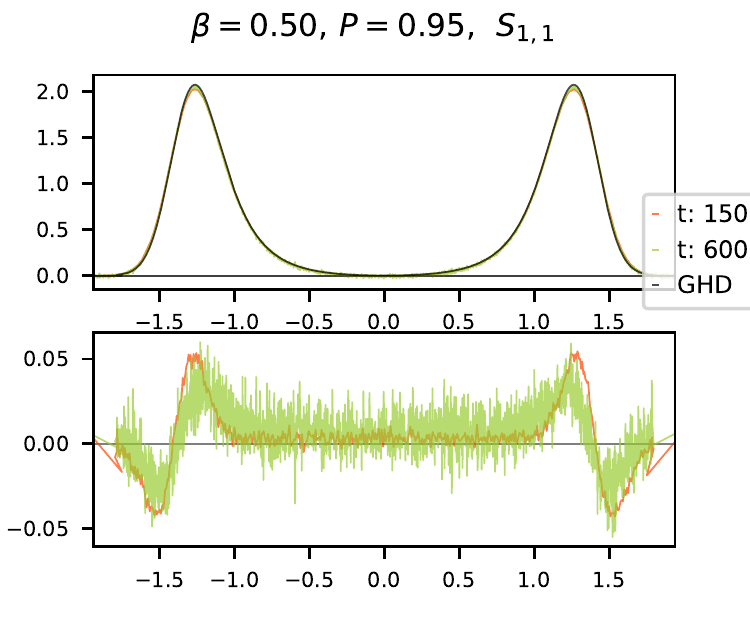}
\includegraphics{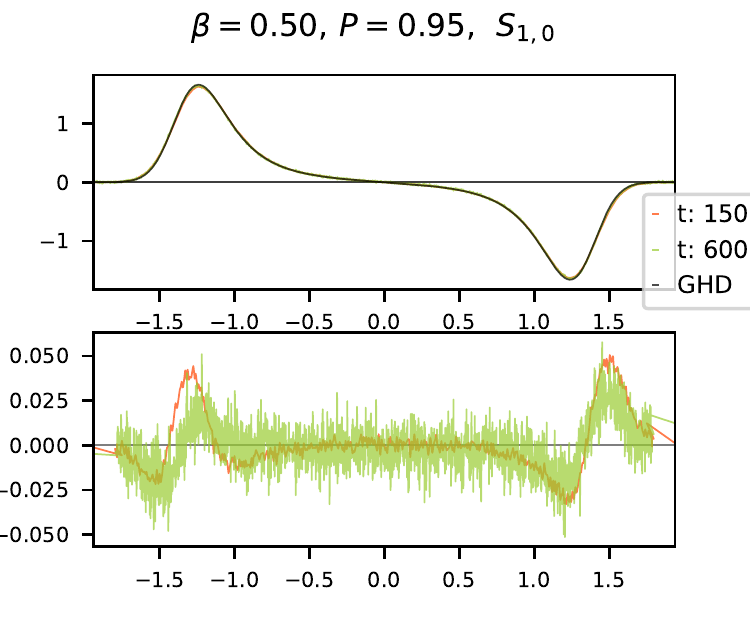}
\caption{Toda correlation functions $S_{1,1}$ (left) and $S_{1,0}$ (right) for medium pressure and increasing temperatures (top to bottom). For each value of $\beta$ and $P$ the top panels show GHD prediction vs. numerical simulations as in Figures~\ref{fig:ontop_beta05}-\ref{fig:ontop_beta20} but with the the molecular dynamics evaluated at two times $t=150$ and $t=600$. The 
bottom panels display the differences between the GHD prediction and numerical simulations at time $t=150$ (red) and at time $t=600$ (green). Number of trials: $3\times10^6$.}
\label{fig:compa_error}
\end{figure}

\begin{figure}
\centering
\includegraphics{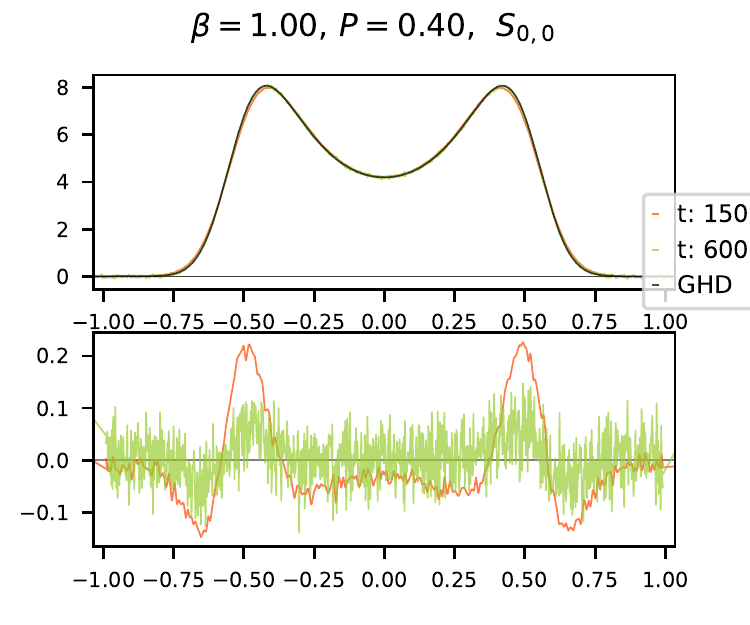}
\includegraphics{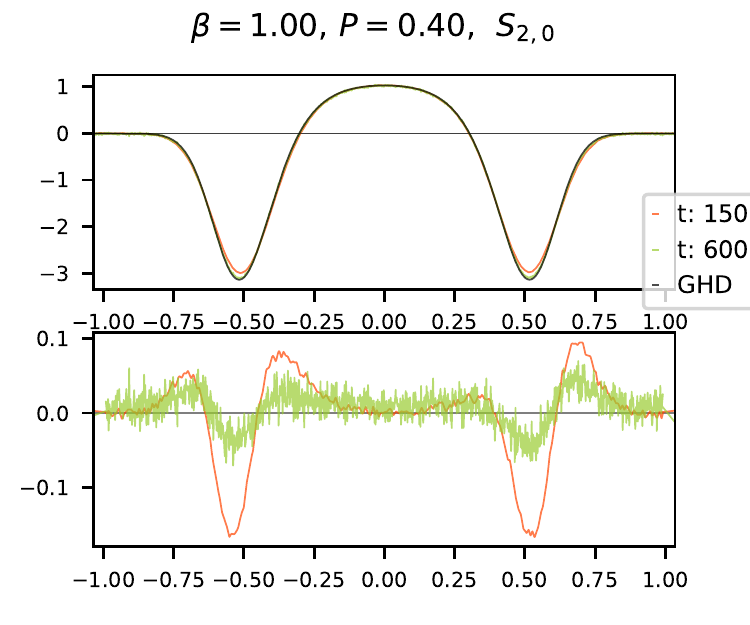} \\
\includegraphics{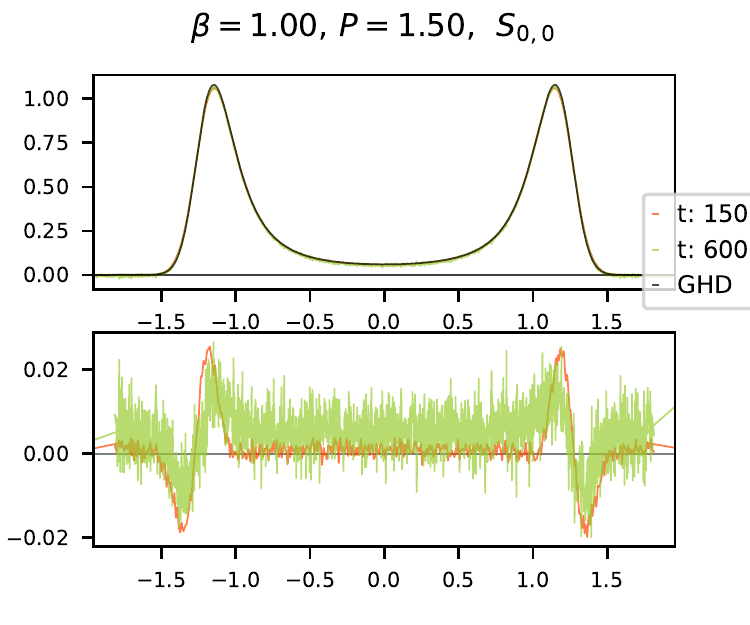}
\includegraphics{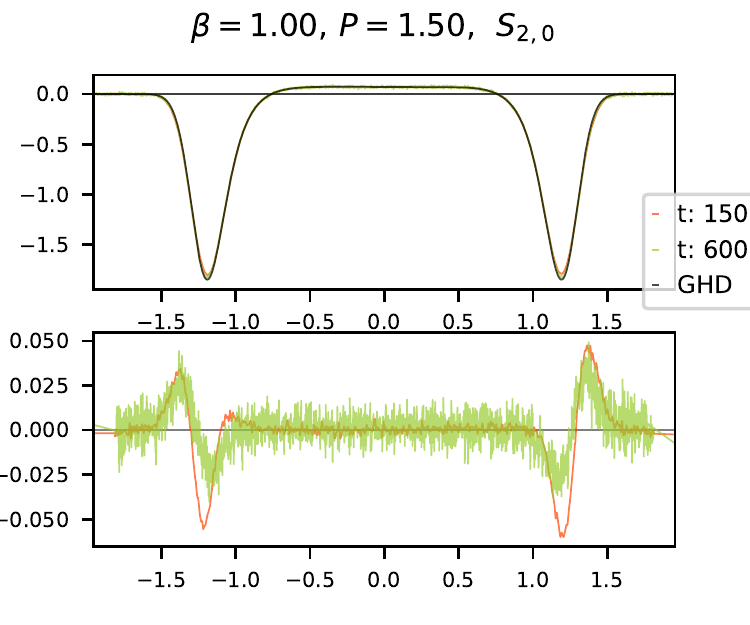}\\
\includegraphics{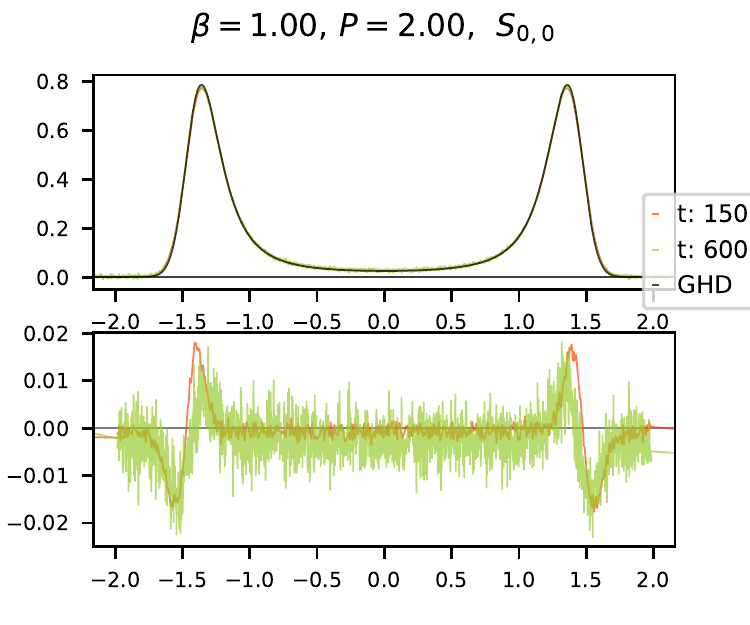}
\includegraphics{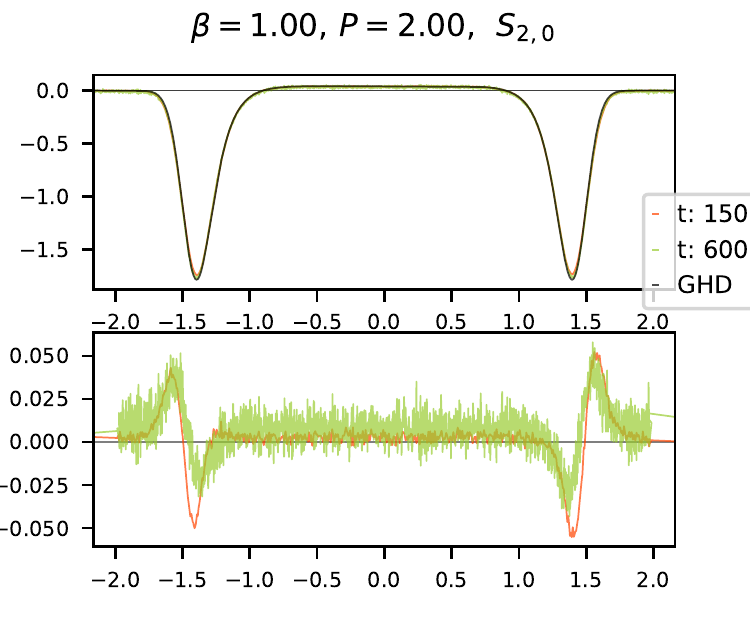}
\caption{Toda correlation functions $S_{0,0}$ (left) and $S_{2,0}$ (right) for $\beta=1$ and increasing pressure (top to bottom). For each value of $\beta$ and $P$ the top panels show GHD prediction vs. numerical simulations as in Figure~\ref{fig:ontop_beta10} but with the the molecular dynamics evaluated at two times $t=150$ and $t=600$. The 
bottom panels display the differences between the GHD prediction and numerical simulations at time $t=150$ (red) and at time $t=600$ (green).Number of trials: $3\times10^6$.}
\label{fig:compa_error2}
\end{figure}

The purpose of this subsection is twofold. On the one hand we have a look at the small differences between GHD predictions and molecular dynamics simulations that can hardly be detected in Figures~\ref{fig:ontop_beta05}-\ref{fig:ontop_beta20}. On the other hand we indicate how these differences evolve in time by including time $t=150$ for the molecular dynamics. Recall that the GHD predictions are time-invariant in the scaling $y \mapsto t S_{m,n}(yt,t)$ we have chosen, see \eqref{48}.

From the 54 comparisons that are displayed in Figures~\ref{fig:ontop_beta05}-\ref{fig:ontop_beta20} we select 12 cases that are representative and show all the phenomena that we have observed. In Figure~\ref{fig:compa_error} we consider correlations $S_{1,1}$ and $S_{1,0}$ at medium pressure (cf.~Table \ref{Table_average_stretches}) for all three values of $\beta$. The small scale fluctuations displayed in the bottom panels are due to the approximation of expectation values by empirical averages. Their amplitudes become smaller if one increases the number of trials. Note that the difference in amplitudes of these fluctions between $t=150$ and $t=600$ is mostly due to the scaling $y \mapsto t S_{m,n}(yt,t)$ that we use. This implies that the values of the correlations are multiplied by a factor that is $4$ times larger at the later time. The same holds for the plots in Figure~\ref{fig:compa_error2} where the correlations $S_{0,0}$ and $S_{2,0}$ are shown for fixed $\beta=1$ and our three different choices for pressure. We now summarize our main findings:
\begin{itemize}
    \item[1.] The deviations occur mostly near the sound peaks and amount to $1.5\%$-$3.5\%$ of the peaks' maximal values at time $t=600$.
    \item[2.] There appear to be small but systematic deviations concerning the shape of the sound peak in all cases. One would need to conduct experiments with a higher resolution, i.e.~more sites and consequently larger times and more trials, to determine whether there is indeed such a systematic deviation. With the resolution present in our experiments the question of a systematic deviation with respect to the shape of the peak cannot be decided.
    \item[3.] In some of the experiments the maximal deviations would be significantly smaller if a constant only depending on the values of $\beta$, $P$, $m$, $n$ is added to all values of $S_{m,n}(j,t)$, see e.g. correlations $S_{0,0}$ and $S_{2,0}$ for $\beta=1$, $P=0.4$ in Figure~\ref{fig:compa_error2}. This seems to be related to the approximation errors for the means $\langle r \rangle$, $\langle p \rangle$, and $\langle e \rangle$, that appear to be less pronounced in the case of momentum $p$. We have observed that these deviations decrease as the number of trials is increased and we do not expect a systematic deviation between GHD and MD in this respect.
    \item[4.] For $(\beta; P) \in \{(0.5; 0.95), (0.5; 1.21)\}$ we observe that the size of the deviations is essentially the same for times $t=150$ and $t=600$ whereas for $(\beta; P) \in \{(0.5; 0.32), (1; 0.4), (2; 0.52),$ $(2; 2.55), (2; 3.53)\}$ these deviations are significantly larger at the smaller time. The remaining two cases $(\beta; P) \in \{(1; 1.5), (1; 2)\}$ are somewhat in between, also depending on the correlation function that is considered, see Figure~\ref{fig:compa_error2}. This is an indication that the speed of convergence of $t S_{m,n}(yt,t)$ to the GHD prediction $S^\mathrm{LL}_{m,n}(y,1)$ as $t \to \infty$ depends on the values of $\beta$ and $P$. As a rule we have observed that both increasing temperature or increasing pressure leads to a faster speed of convergence.
\end{itemize}

\section{Conclusions and outlook}
\label{sec:outlook}
As can be seen from Table 1, we picked the intermediate pressure such that $\nu \simeq 0$. 
In the particle picture $\nu = 0$ corresponds to the boundary condition $q_1 = q_N$. In thermal equilibrium
the positions then perform an unbiased random walk with typical excursions of order $\sqrt{N}$. Thus the free volume is of order
$1/\sqrt{N}$. The particles are extremely dense and the picture of successive pair collisions breaks down completely.
So one might wonder whether GHD is still valid under such extreme conditions. $\nu =0$ poses no particular 
difficulties for MD simulations. In GHD the factor $1/\nu$ appears  in the expression for $v^\mathrm{eff}$, see Eq. \eqref{47}. This makes the numerical scheme slow and only values close to $\nu =0$ are accessible. However the correlator
$S$ changes smoothly through  $\nu = 0$. GHD also covers this seemingly singular point.

Simultaneously A. Kundu \cite{K22} posted a somewhat puzzling note. He considers the parameter values
$\beta = 1$, $P=1$. When cutting the matrices $C_{m,n}$ and $A_{m,n}$ at low orders, the resulting $S_{m,n}$ consists of  a few $\delta$-peaks which move at constant velocity. After ballistic scaling,  with high precission they turn out to lie on the curve obtained from GHD. A theoretical explanations seems to be missing.

In \cite{KD16} the molecular dynamics of Toda lattice correlations are simulated for the potential
\begin{equation}
V_\mathrm{kd}(x) = \frac{g}{\gamma}\mathrm{e}^{-\gamma x}
\end{equation}
with arbitrary $\gamma, g > 0$. To distinguish their parameters from ours, the variables in \cite{KD16}  are here denoted by $\bar{t},\bar{r}, \bar{P},\bar{\beta}$. $\bar{P}$ is the physical pressure and, comparing the Gibbs weights, one obtains the relations
\begin{equation} 
\beta = \frac{g}{\gamma}\bar{\beta},\qquad P = \frac{1}{\gamma} \bar{P}\bar{\beta}.
\end{equation}
From the equations of motions one deduces
\begin{equation}\label{52} 
\bar{t} = \frac{1}{\sqrt{\gamma g}} t, \quad r(t) = \gamma\bar{r}(\bar{t}),\quad p(t) = \frac{g}{\gamma}\bar{p}(\bar{t}).
\end{equation}
Thus, translating to our units, the MD simulations reported in \cite{KD16}  are (i) $P= 0.01$, $\beta = 0.01$, $N=1024$, $t = 400$, (ii) $P= 1$, $\beta = 1$, $N=1024$, $t = 200,300$, and  (iii)
$P= 400$, $\beta = 400$, $N= 256$, $t = 80$.  In fact, in all three cases the time scales are identical, $t = \bar{t}$.  Since GHD was not available
yet, no comparison could have been attempted. 

Case (i) is a very dilute chain. In this limit $\nu \rho_\mathsf{p}$ is a unit Gaussian. The dressed functions become polynomials as $\varsigma_0^\mathrm{dr}(w) = a_0$, $\varsigma_1^\mathrm{dr}(w) = a_1 w$, and $\varsigma_2^\mathrm{dr}(w) = a_2w^2 + a_3$ with coefficients $a_0,...,a_3$ depending on $(P,\beta)$. 
Note that for a noninteracting fluid $a_3$ would vanish.
As a result $S_{0,0}$ is Gaussian,  $S_{1,1}$ has two peaks, and $S_{2,2}$ has either two or three peaks. 
This is in good agreement with \cite{KD16} and explains our motivation not to venture into the low density regime.
Case (ii) interpolates between our $\beta = 1,P = 0.40$ and $\beta = 1,P = 1.5$. Note that now $S_{0,0}$ has a local minimum at $w=0$,
which is very different from the structure in the dilute regime. On the other hand, $S_{2,2}$ has a local maximum at $w=0$,
as is the case for low density/high temperature.

The most interesting parameter value is (iii), which deserves more detailed studies. The issue is the behavior of the
Toda chain at very low temperatures. Simply letting $\beta \to \infty$ will freeze any motion. But the simultaneous limit 
$\beta \to \infty$ with $P = \bar{P}\beta$ at fixed physical pressure $\bar{P}$  is meaningful, at least statistically. In this limit $\nu > 0$ always.  
 Also the density of states converges to the arcsine distribution,
 \begin{equation} 
\lim_{\beta \to \infty} \nu \rho_\mathsf{p}(w) = \frac{1}{\pi \sqrt{4\bar{P} -w^2}}, \quad |w| \leq 2\sqrt{\bar{P}}.
\end{equation}
To understand the dynamical behavior, the effective potential is expanded as
 \begin{equation} 
\mathrm{e}^{-r} + \bar{P}r \simeq \tfrac{1}{2}\bar{P}(r-r_0)^2 +c_0
\end{equation}
at its minimum $r_0$. Since $\beta$ is large, the initial fluctuations are of order $1/\sqrt{\beta}$. Therefore the dynamics
can be approximated by a harmonic chain with $\omega^2 =  \bar{P}$. The equilibrium time correlations of the harmonic chain have intricate oscillatory behavior \cite{GKMM21}, which in the large $\beta$ limit should match with the Toda lattice,
as partially evidenced through case (iii). Clearly, GHD cannot reproduce such  fine details. Still, when averaged on suitable
scales, the gross behavior of the harmonic chain oscillations might be visible. 

\section*{Acknowledgements}
This material is based upon work supported by the National Science Foundation under Grant No. 1440140, while five of the authors were in residence at the Mathematical Sciences Research Institute in Berkeley, California, during the fall semester of 2021.

The authors would like to thank the Isaac Newton Institute for Mathematical Sciences, Cambridge, for support and hospitality during the programme ``Dispersive hydrodynamics: mathematics, simulation and experiments, with applications in nonlinear waves" where some work on this paper was undertaken. This work was supported by EPSRC grant no EP/R014604/1. TG acknowledges the support of  the European Union's H2020   Marie Sk\l odowska--Curie grant No. 778010 {\em  IPaDEGAN},  of INdAM/GNFM and  of  the research project Mathematical Methods in NonLinear Physics (MMNLP), Gruppo 4-Fisica Teorica of INFN. GM is financed by the KAM grant number 2018.0344.  KTRM was supported by a Visiting Wolfson research fellowship from the Royal Society.

\bibliographystyle{siam}
\bibliography{newbib}

\end{document}